\documentstyle[seceq,preprint]{ptptex}

\newcommand{\tr}{{\rm tr}}
\newcommand{\Tr}{{\rm Tr}}

\newcommand{\citea}[2]{#2~\cite{#1} }

\preprintnumber{
CHIBA-EP-104\\ hep-th/9803063}

\title{
Abelian-Projected Effective Gauge Theory of QCD 
with Asymptotic Freedom and Quark Confinement 
\footnote{Invited talk given at YKIS'97, Non-perturbative QCD: 
 Structure of the QCD vacuum, held at YITP,
Kyoto University, 2-12 December 1997,
to appear in Progress of Theoretical Physics Supplement.} 
}

\author{
Kei-Ichi {\sc Kondo}\footnote{E-mail:  kondo@cuphd.nd.chiba-u.ac.jp} 
}

\inst{
Department of Physics, Faculty of Science,
  Chiba University, Chiba 263-8522, Japan
}

\recdate{
\today
}

\abst{
We give an outline of a recent proof
that the low-energy effective gauge theory exhibiting quark
confinement due to magnetic monopole condensation can be derived
from QCD without any specific assumption. We emphasize that the
low-energy effective abelian gauge theories obtained here give the
dual description of the same physics in the low-energy region.    
They show that the QCD vacuum is nothing but the dual (type II)
superconductor. 
}

\begin{document}

\maketitle

\section{Introduction}
\par
The ideal of dual superconductor vacuum of
QCD as an origin of quark confinement was proposed by
Nambu,\cite{Nambu74} Mandelstam
\cite{Mandelstam76} and 't Hooft in the mid-1970.  
Although there may be various scenarios for quark confinement, the
picture of dual superconductor vacuum of QCD is the most intuitively
appealing one, since this is based on the dual picture of
the well-known ordinary superconductivity.
In the ordinary superconductivity due to  Cooper pair condensation,
 magnetic flux is squeezed into the tube-like region in the type
II superconductor.  
Here the term 'dual' implies that the role of electric
field and magnetic field is interchanged.  
Hence it is expected that, if the magnetic monopole condensates, 
the color electric flux is squeezed into the string connecting
the quark and anti-quark pair.  This leads to the linear static
potential. However, why the dual superconductor appears from QCD was
a mystery before a resolution was proposed by 't Hooft
\cite{tHooft81} in 1981.
\par
\underline{Abelian projection}:
The basic idea of abelian projection proposed by 't Hooft
\cite{tHooft81} is to remove as many non-Abelian degrees of freedom
as possible, by partially fixing the gauge in such a way that {\it
maximal torus group} $H$ of the gauge group $G$ remains unbroken.
For $G=SU(N)$, $H=U(1)^{N-1}$.  He
claimed that under the abelian projection the
$SU(N)$ gauge theory reduces to an $U(1)^{N-1}$ abelian gauge theory
plus {\it magnetic monopole}.
\par
\underline{Abelian dominance hypothesis}:
Soon after the proposal of abelian projection, the hypothesis of
abelian dominance was proposed by Ezawa and Iwazaki \citea{EI82}. 
The abelian dominance claims that the non-Abelian components do  not
contribute to the physics in the low energy or at a long-distance
scale.  This was a hypothesis at that time.  However, abelian
dominance has been confirmed based on Monte Carlo simulation on the
lattice by Suzuki and Yotsuyanagi \cite{SY90} in 1990.
\par
\underline{Massive off-diagonal gluons}
Why is the abelian dominance realized in QCD?
A basic observation is as follows.
Abelian dominance will be achieved if the QCD dynamics makes the
non-Abelian component {\it heavier} than the Abelian component, so
that the non-Abelian components do not propagate at long-distance
scale. 
\par
\underline{Wilsonian RG}:
How to prove abelian dominance?
We need to derive the effective gauge theory of YM theory at
resolution (length scale) $R$ where all the field variables with
momentum $p \ge R^{-1}$ are integrated out in the sense of Wilsonian
renormalization group (RG).
\par
\underline{QCD at scale $R$}
In the problem of quark confinement, $R$ could be the distance
between two quarks.  If $R$ is small, the perturbative picture is
valid ($\Longrightarrow$ Asymptotic freedom).
As $R$ increases, the perturbative picture gradually becomes
dubious.  When $R$ reaches at a certain critical distance $R_c$,
monopole condensation is expected to occur.
For $R>R_c$, electric vortices emerge as stable topological
excitations and confine quarks ($\Longrightarrow$ confinement).
\par
Starting from SU(2) Yang-Mills theory in 3+1 dimensions,
we prove that the abelian-projected effective gauge
theories are written in terms of the maximal abelian gauge
field and the dual abelian gauge field interacting with
monopole current.  This is performed by integrating out
all the remaining non-Abelian gauge field belonging to
SU(2)/U(1).  We show that the resulting abelian gauge
theory recovers exactly the same one-loop beta function as
the original Yang-Mills theory.  Moreover, the dual abelian
gauge field becomes massive if the monopole condensation
occurs.  This result supports the dual superconductor
scenario for quark confinement in QCD.  We give a
criterion of dual superconductivity. Therefore there can
exist the effective abelian gauge theory which shows both
asymptotic freedom and quark confinement based on the dual
Meissner mechanism.
\par
In the following we give an outline of the proof.
For the details and references, see  the original paper with the
same title
\citea{Kondo97}.
\par
For more backgrounds, see contributions of
A. van der Sijs, A. Di Giacomo, H. Toki, M. Polikarpov, 
H. Markum, H. Suganuma, T. De Grand, O. Miyamura, T. Suzuki and K.
Schilling.

\section{Basic strategy}

\par
How to derive the low energy effective (maximal abelian)
gauge theory (LEEGT)  for a given non-Abelian gauge theory with a
non-Abelian gauge group $G$.  A strategy is as follows.
\begin{enumerate}
\item[Step 1:]  Cartan decomposition,
\begin{eqnarray}
 G = H \otimes G/H, \quad {\cal A} = \{ a, A \} 
 \in {\cal H} \oplus {\cal G}-{\cal H} .
\end{eqnarray}
\item[Step 2:]  Gauge fixing to fix off-diagonal parts (abelian
gauge),
\begin{eqnarray}
  A_\mu = A_\mu^a T^a \in {\cal G}-{\cal H} .
\end{eqnarray}

\item[Step 3:]  Integrate out off-diagonal non-Abelian parts.

\end{enumerate}

Then we obtain the  effective gauge theory 
written in terms of
$a_\mu \in {\cal H}$,
\begin{eqnarray}
 \int [da_\mu] \int [dA_\mu] e^{iS_{YM}[a,A]} \delta(F[A])
 = \int [da_\mu] e^{iS_{LEEGT}[a]} .
\end{eqnarray}
However, this scenario has the following difficulties.
\par
Q1: Where does the magnetic monopole come from?
\par
Q2: How can we integrate out off-diagonal gluon fields which have
quartic self-interaction?
\par
Both difficulties are resolved by adopting an alternative strategy.
\par
Step 1': Introduce auxiliary (antisymmetric) abelian tensor field
$B_{\mu\nu}$.  This enables us to perform exact integration of the
off-diagonal gluon fields, 
$A_\mu = A_\mu^a T^a \in {\cal G}-{\cal H}$.  Then we obtain 
the abelian-projected effective gauge theory (APEGT) as a LEEGT of
QCD,
\begin{eqnarray}
 \int [da_\mu] \int [dA_\mu] e^{iS_{YM}[a,A]} \delta(F[A])
 &=& \int [da_\mu] \int [dB_{\mu\nu}]
 \int [dA_\mu] e^{iS_{YM}[a,A,B]} \delta(F[A])
 \nonumber\\
 &=& \int [da_\mu] \int [dB_{\mu\nu}] e^{iS_{APEGT}[a,B]} .
\end{eqnarray}

\section{Derivation of APEGT}
\subsection{Step 1': Cartan decomposition}
Perform the Cartan decomposition,
\begin{eqnarray}
 {\cal A}_\mu(x) = \sum_{A=1}^3  {\cal A}_\mu^A(x) T^A
 :=  a_\mu(x)  T^3 
 + \sum_{a=1}^{2}  A_\mu^a(x) T^a 
  \in {\cal H} \oplus {\cal G}-{\cal H} .
\end{eqnarray}
Then the field strength 
\begin{eqnarray}
 {\cal F}_{\mu\nu}(x) 
:= \sum_{A=1}^{3} {\cal F}_{\mu\nu}^A(x) T^A
:=   \partial_\mu {\cal A}_\nu(x) 
 -   \partial_\nu {\cal A}_\mu(x)
 - i [{\cal A}_\mu(x), {\cal A}_\nu(x)]
\end{eqnarray}
is decomposed as 
\begin{eqnarray}
 {\cal F}_{\mu\nu}(x) 
 &=&  [f_{\mu\nu}(x) + {\cal C}_{\mu\nu}(x)]T^3 
 + {\cal S}_{\mu\nu}^a(x)T^a ,
\nonumber\\
 f_{\mu\nu}(x) &:=& \partial_\mu a_\nu(x) 
 -   \partial_\nu a_\mu(x) ,
\nonumber\\
 {\cal S}_{\mu\nu}^a(x) &:=& 
 D_\mu[a]^{ab} A_\nu^b - D_\nu[a]^{ab}A_\mu^b ,
\nonumber\\
 {\cal C}_{\mu\nu}(x)T^3 &:=& - i[ A_\mu(x), A_\nu(x)] ,
\end{eqnarray}
where the derivative $D_\mu[a]$ is defined by
\begin{eqnarray}
 D_\mu[a]  = \partial_\mu   + i[a_\mu T^3, \cdot ~],
\quad
 D_\mu[a]^{ab} := \partial_\mu \delta^{ab}
 - \epsilon^{ab3} a_\mu .
\end{eqnarray}
Next, we rewrite the Yang-Mills (YM) action
\begin{eqnarray}
   S_{YM}[{\cal A}] &=& -{1 \over 2g^2} \int d^4x \
   \tr({\cal F}_{\mu\nu} {\cal F}^{\mu\nu}) 
   \nonumber\\
   &=& -{1 \over 4g^2} \int d^4x \
   [(f_{\mu\nu} + {\cal C}_{\mu\nu})^2
    + ({\cal S}_{\mu\nu}^a)^2] .
\label{YM2}
\end{eqnarray}
We wish to integrate out the off-diagonal gluons.  
Note that ${\cal C}_{\mu\nu}$ is quadratic  in $A$.  Hence
${\cal C}_{\mu\nu}^2$ is quartic in $A$.  All the other terms are
integrated out by Gaussian integration.
In order to perform Gaussian integration for all the terms, we
introduce auxiliary antisymmetric tensor field $B_{\mu\nu}$.  Note
that
$B_{\mu\nu}$ is an abelian tensor field.  We can consider two
equivalent theories depending on the way we introduce the auxiliary
tensor field.
\par
(I)
If we introduce the auxiliary tensor field in such a way
\begin{eqnarray}
   -{1 \over 4g^2}(f_{\mu\nu} + {\cal C}_{\mu\nu})^2 \rightarrow 
     {1 \over 4} \epsilon^{\mu\nu\rho\sigma} B_{\rho\sigma} 
     (f_{\mu\nu} + {\cal C}_{\mu\nu}) 
     - {1 \over 4} g^2 B_{\mu\nu} B^{\mu\nu} ,
    \label{apBFYM0}
\end{eqnarray}
which corresponds to the tree-level duality,
$
  B_{\mu\nu}  \leftrightarrow {1 \over 2}
\epsilon^{\mu\nu\rho\sigma} {\cal F}^3_{\rho\sigma}
=  {1 \over 2} \epsilon^{\mu\nu\rho\sigma}
(f_{\rho\sigma} + {\cal C}_{\rho\sigma}) ,
$
we obtain
\begin{eqnarray}
   S[{\cal A}, B] =  \int d^4x \left[
     {1 \over 4} \epsilon^{\mu\nu\rho\sigma} B_{\rho\sigma} 
     (f_{\mu\nu} + {\cal C}_{\mu\nu}) 
     - {1 \over 4} g^2 B_{\mu\nu} B^{\mu\nu} 
    - {1 \over 4g^2} ({\cal S}_{\mu\nu}^a)^2 \right] .
    \label{apBFYM}
\end{eqnarray}
This theory is equivalent to the BF-YM theory, i.e. a deformation of
the topological BF theory, see Ref.~\citen{Kondo97}.

\par
(II)
If we take
\begin{eqnarray}
    -{1 \over 4g^2}({\cal C}_{\mu\nu})^2 \rightarrow 
   {1 \over 4} \epsilon^{\mu\nu\rho\sigma} B_{\rho\sigma} 
   {\cal C}_{\mu\nu}
     - {1 \over 4} g^2 B_{\mu\nu} B^{\mu\nu} ,
    \label{apYM0}
\end{eqnarray}
which corresponds to 
$
  B_{\mu\nu}  \leftrightarrow   {1 \over 2}
\epsilon^{\mu\nu\rho\sigma} {\cal C}_{\rho\sigma} ,
$
we are lead to the action,
\begin{eqnarray}
   S[{\cal A}, B] &=&  \int d^4x \Big[
     - {1 \over 4g^2} 
     (f_{\mu\nu}f_{\mu\nu} + 2 f_{\mu\nu}{\cal C}_{\mu\nu}) 
  +  {1 \over 4} \epsilon^{\mu\nu\rho\sigma} B_{\rho\sigma} 
   {\cal C}_{\mu\nu}
     - {1 \over 4} g^2 B_{\mu\nu} B^{\mu\nu} 
\nonumber\\&&
    - {1 \over 4g^2} ({\cal S}_{\mu\nu}^a)^2 \Big] .
    \label{apYM}
\end{eqnarray}

\par
We can show that two theories are equivalent to each other, see
Ref.~\citen{Kondo97}.
In what follows, we focus on the action
(\ref{apYM}).

\subsection{Step 2: Maximal abelian gauge fixing}
We consider the gauge fixing,
\begin{eqnarray}
 F^{\pm}[A,a] &:=& (\partial^\mu \pm i \xi a^\mu)
A_\mu^{\pm} = 0,
\label{dMAG}
\\
  F^3[a] &:=& \partial^\mu  a_\mu = 0 ,
 \label{GF}
\end{eqnarray}
where we have used the $(\pm, 3)$ basis.
\footnote{
$
{\cal O}^{\pm} := ({\cal O}^1 \pm i{\cal O}^2)/\sqrt{2} .
$
In this basis, 
$
 \sum_{\pm} P^{\pm}Q^{\mp} = P^+ Q^- + P^- Q^+ = P^a Q^a ,
$
and
$
 \sum_{\pm} (\pm) P^{\mp}Q^{\pm} = - P^+ Q^- + P^- Q^+ =  i
\epsilon^{ab3} P^a Q^b (a,b=1,2).
$
}
The gauge fixing with $\xi=0$ is the Lorentz gauge,
$\partial^\mu {\cal A}^\mu = 0$.
In particular, $\xi=1$ corresponds to the differential form
of the maximal abelian gauge (MAG) which is expressed as
the minimization of the functional 
\begin{eqnarray}
  {\cal R}[A] := {1 \over 2}\int d^4x [(A_\mu^1(x))^2 +
(A_\mu^2(x))^2 ]
  = \int d^4x A_\mu^{+}(x) A_\mu^{-}(x)   .
  \label{MAG}
\end{eqnarray}
The differential MAG condition (\ref{dMAG}) corresponds to
 a local minimum of the gauge fixing functional
${\cal R}[A]$, while the MAG condition (\ref{MAG}) requires the
global (absolute) minimum.  
The differential MAG condition (\ref{dMAG}) fixes gauge
degrees of freedom in SU(2)/U(1) and is invariant under the
residual U(1) gauge transformation.  An additional
condition (\ref{GF}) fixes the residual U(1) invariance. 
Both conditions (\ref{dMAG}) and 
(\ref{GF}) then completely fix the gauge except
possibly for the Gribov problem.
It is known that the differential MAG
(\ref{dMAG}) does not spoil renormalizability of YM theory. 

\par
From physical point of view, we expect that MAG introduces
the non-zero mass $m_{A}$ for the off-diagonal gluons,
$A_\mu^1, A_\mu^2$.
This is suggested from the form (\ref{MAG}) which is
equal to the mass term for $A_\mu^1, A_\mu^2$, although
we need an independent proof of this statement.
\footnote{
This statement has been proved soon after this talk, see
Ref.~\citen{Kondo98}. }
This motivates us to integrate out the off-diagonal gluons in
the sense of Wilsonian renormalization group and allows us
to regard the resulting theory as the low-energy effective
gauge theory written in terms of massless fields alone
which describes the physics in the length scale 
$R > m_{A}^{-1}$.
The abelian dominance will be realized in the physical
phenomena occurring in the scale $R > m_{A}^{-1}$. 

\par
We introduce the Lagrange multiplier field
$\phi^{\pm}$ and $\phi^3$ for the gauge-fixing
function $F^{\pm}[A]$ and
$F^{3}[A]$, respectively. 
It is well known that the gauge fixing term and the Faddeev-Popov
ghost term are obtained using the BRST transformation $\delta_B$
as
\begin{eqnarray}
  {\cal L}_{GF} = - i \delta_B G ,
    \label{BRSTGF}
\end{eqnarray}
where $G$ carries the ghost number $-1$ and is a hermitian
function of Lagrange multiplier field $\phi^{\pm}, \phi^3$,
ghost field $c^A$, antighost field $\bar c^A$, and the
remaining field variables of the original lagrangian.  
The simplest choice is given by
\begin{eqnarray}
  G = \sum_{\pm}
  \bar C^{\mp} (F^{\pm}[A,a] + {\alpha \over 2} \phi^{\pm})
  + \bar C^3 (F^{3}[a] + {\beta \over 2} \phi^{3}) .
\label{G1}
\end{eqnarray}
\par
Under the local U(1) gauge transformation,
\begin{eqnarray}
 a_\mu \rightarrow a_\mu + \partial_\mu \omega, 
 \quad
 {\cal O}^{\pm} \rightarrow e^{\mp i \omega} {\cal O}^{\pm}
 \quad
 {\cal O}^{3} \rightarrow {\cal O}^{3} .
\end{eqnarray}
Hence $a_\mu$ transforms as a U(1) gauge field, while
 $A_\mu^{\pm}$  and $B_{\mu\nu}^{\pm}$  behave as
charged matter fields under the U(1) gauge transformation.
It turns out that
$B_{\mu\nu}^3$ and 
$
 {\cal C}_{\mu\nu} 
= i \sum_{\pm} (\pm) A_\mu^{\pm} A_\nu^{\mp}
$
are U(1) gauge invariant as expected. 
\par
\par
For the gauge fixing function (\ref{G1}), straightforward
calculation leads to the gauge fixing lagrangian (\ref{BRSTGF}),
\begin{eqnarray}
  {\cal L}_{GF} &=& 
  \phi^a F^a[A,a] + {\alpha \over 2} (\phi^a)^2
  + i \bar C^a D^\mu{}^{ab}[a]^\xi D_\mu^{bc}[a] C^c
  \nonumber\\
  &&- i \xi \bar C^a [A_\mu^a A^\mu{}^b - A_\mu^c A^\mu{}^c
\delta^{ab} ] C^b
  \nonumber\\
  && + \phi^3 F^3[a] + {\beta \over 2} (\phi^3)^2
  + i \bar C^3 \partial^\mu \partial_\mu C^3
  - i  \bar C^3 \partial^\mu (\epsilon^{ab3} A_\mu^a C^b)  
  \nonumber\\
  &&+ i \bar C^a \epsilon^{ab3} [(1-\xi) A_\mu^b \partial^\mu
  + F^b[A,a] ] C^3 .
    \label{BRSTGF2}
\end{eqnarray}
This reduces to the usual form in the Lorentz gauge,
$\xi=0$.
\par

\subsection{Step 3: Integration over SU(2)/U(1)}
We perform integration over the off-diagonal fields 
$\phi^a, A_\mu^a, C^a, \bar C^a$ (and $B_{\mu\nu}^a$ for BF-YM case)
belonging to the Lie algebra of SU(2)/U(1) and obtain the
effective abelian gauge theory written in terms of the diagonal
fields
$a_\mu, B_{\mu\nu}$. 
\par
Under the MAG, the total action is given by
\begin{eqnarray}
   S_{YM} &=& S_{YM}[a,A,B,C,\bar C] =
   S_1[a,B] + S_2[a,C,\bar C] + S_0[a,B,C,\bar C] ,
    \label{apBFYM3}
\\
   S_1 &=&  \int d^4x \left[
    - {1 \over 4g^2}  f_{\mu\nu} f^{\mu\nu}
    - {1 \over 4} g^2 B_{\mu\nu} B^{\mu\nu}
\right]  ,
\label{S1}
\\
 S_2 &=&  \int d^4x \left[
 i \bar C^a D^\mu{}^{ac}[a] D_\mu^{cb}[a] C^b
+ i \bar C^3 \partial^\mu \partial_\mu C^3
+  \phi^3 (\partial^\mu a_\mu) + {\beta \over 2} (\phi^3)^2
\right]  ,
\label{S2}
\\
    S_0
&=& -i \ln \int [dA_\mu^a] \exp \left\{ i \int d^4x \left[
  {1 \over 2g^2} A_\mu^a  Q_{\mu\nu}^{ab} A_\nu^b  
+ A_\mu^a  G_\mu^a  
 \right] \right\}
\nonumber\\
&=& - {1 \over 2} \ln \det (Q_{\mu\nu}^{ab}) 
 + {g^2 \over 2}  G_\mu^a (Q^{-1})_{\mu\nu}^{ab} G_\nu^b ,
\label{S3}
   \\
     Q_{\mu\nu}^{ab} &:=&  
   (D_\rho[a] D_\rho[a])^{ab} \delta_{\mu\nu}
   - 2 \epsilon^{ab3} f_{\mu\nu}
 + {1 \over 2} g^2 \epsilon^{ab3}
   \epsilon_{\mu\nu\rho\sigma} B^{\rho\sigma} 
\nonumber\\&&
 - 2i g^2  (\bar C^a  C^b - \bar C^c C^c \delta^{ab})
\delta_{\mu\nu}
 - D_\mu[a]^{ac} D_\nu[a]^{cb} 
 + {1 \over \alpha}  D_\mu[a]^{ac} D_\nu[a]^{cb} ,
\label{defK}
\\
 G_\mu^c  &:=& i (\partial_\mu \bar C^3 )
\epsilon^{cb3} C^b - i \bar C^a \epsilon^{ab3} \epsilon^{bc3}
a_\mu C^3  - i \partial_\mu(\bar C^a \epsilon^{ac3} C^3) .
    \label{defG}
\end{eqnarray}
where we have rescaled the parameter $\alpha$ to absorb the
$g$ dependence.
\par
The residual U(1) invariant theory is obtained by putting 
$\phi^3 = 0$ and $\bar C^3 = C^3 = 0$ (hence
$G_\mu^a = 0$).  Therefore, the resulting APEGT is greatly
simplified,
\par
\begin{eqnarray}
   S_{YM} &=& \int d^4x \left[
    - {1 \over 4g^2}  f_{\mu\nu} f^{\mu\nu}
    - {1 \over 4} g^2 B_{\mu\nu} B^{\mu\nu}
 + i \bar C^a D^\mu{}^{ac}[a] D_\mu^{cb}[a] C^b \right]
 \nonumber\\&&
 - {1 \over 2} \ln \det (Q_{\mu\nu}^{ab}) .
\label{S}
\end{eqnarray}
The logarithmic determinant is calculated using the $\zeta$-function
regularization.
\begin{eqnarray}
  \ln \det Q = - \lim_{s \rightarrow 0} {d \over ds}
{\mu^{2s} \over \Gamma(s)} \int_{0}^{\infty} dt \ t^{s-1}
\Tr(e^{-t Q}) .
\label{hk}
\end{eqnarray}
This is a gauge-invariant regularization.  Hence the result respects
the residual U(1) gauge invariance.  The result is
\begin{eqnarray}
 {1 \over 2} \ln \det  Q_{\mu\nu}^{ab} 
  &=&   \int d^4x \Big[
  {1 \over 4g^2} z_a f_{\mu\nu} f^{\mu\nu}
    + {1 \over 4} z_b g^2 B_{\mu\nu} B^{\mu\nu}
    +  {1 \over 2} z_c  
B_{\mu\nu} \tilde f_{\mu\nu} 
\nonumber\\&&
\quad + {\rm ghost~self-interaction~terms}
\nonumber\\&&
\quad + {\rm higher~derivative~terms} \Big] ,
\end{eqnarray}
where 
\begin{eqnarray}
 z_a = - {20 \over 3} \kappa {g^2 \over 16\pi^2} \ln \mu , 
\quad
 z_b = + 2 \kappa {g^2 \over 16\pi^2} \ln \mu ,
\quad
 z_c = + 4 \kappa {g^2 \over 16\pi^2} \ln \mu .
 \label{z}
\end{eqnarray}
If we neglect the ghost self-interaction terms and higher derivative
terms, off-diagonal ghost and anti-ghost fields can be integrated
out,
\begin{eqnarray}
S_c &=&  \ln \int [d\bar C][dC] \exp \left\{ - \int d^4x \
\bar C^a D_\mu^{ac}[a] D_\mu^{cb}[a] C^b
\right\}
\nonumber\\
&=& \ln \det (D_\mu^{ac}[a] D_\mu^{cb}[a]) 
\nonumber\\
&=&  \int d^4x {1 \over 4g^2} z_a' f_{\mu\nu} f^{\mu\nu}
+ \cdots ,
\quad
z_a' :=  {2 \over 3} \kappa {g^2 \over 16\pi^2} \ln \mu .
\label{lndetc}
\end{eqnarray}
\par
Finally we obtain 
\begin{eqnarray}
  S_{APEGT}[a,B] 
=
  \int d^4x \left[
    - {Z_a \over 4g^2} f_{\mu\nu} f^{\mu\nu}
    - {1+z_b \over 4} g^2 B_{\mu\nu} B^{\mu\nu}
    +  {1 \over 2} z_c  
B_{\mu\nu} \tilde f_{\mu\nu}   
   \right]  ,
\end{eqnarray}
where 
\begin{eqnarray}
Z_a := 1-z_a+z_a' =
1 + {22 \over 3} \kappa {g^2 \over 16\pi^2} \ln \mu . 
\end{eqnarray}

\section{Asymptotic freedom}

Defining the running coupling constant,
\begin{eqnarray}
 g(\mu) = Z_a^{1/2} g  ,
\end{eqnarray}
the $\beta$-function is easily calculated:
\begin{eqnarray}
  \beta(g) := \mu {dg(\mu) \over d\mu} 
= - {b_0 \over 16\pi^2} g(\mu)^3, 
\quad  b_0 = {11C_2(G) \over 3} > 0 .
\end{eqnarray}
Thus the APEGT exhibits asymptotic
freedom in the sense that APEGT has exactly the same beta function 
as the original YM theory (at one-loop).
In other words, the APEGT is the abelian gauge theory with
QCD-like running coupling constant $g(\mu)$,
\begin{eqnarray}
  S_{APEGT}[a,B]
  &=&  \int d^4x \left[
    - {1 \over 4g(\mu)^2}  f_{\mu\nu}  f^{\mu\nu} 
    + {\rm (B_{\mu\nu}-dependent~terms)}
  \right] ,
 \\
  {1 \over g(\mu)^2} &=&  {1 \over g(\mu_0)^2}
  + {b_0 \over 8\pi^2} \ln {\mu \over \mu_0} ,
  \label{action1}
\end{eqnarray}
apart from the $B_{\mu\nu}$-dependent terms.
\par
If $B_{\mu\nu}$-dependent terms are absent, the kinetic term for the
field
$a_\mu$ does not change due to higher order expansion and hence the
beta function is unchanged, namely, the same as the one-loop result.
Therefore, $B_{\mu\nu}$-dependent terms can give non-perturbative
contribution.

\section{APEGT with magnetic monopole current}

We show that APEGT contains magnetic monopole.
The two-form $B_{\mu\nu}$ has the Hodge decomposition in four
dimensions,
\begin{eqnarray}
 B^{(2)} = db^{(1)} + \delta C^{(3)}
 = db^{(1)} + \delta * \chi ^{(1)}
 = db^{(1)} +  * d \chi^{(1)} .
\end{eqnarray}
This is equivalent to
\begin{eqnarray}
  B_{\mu\nu} = b_{\mu\nu} + \tilde \chi_{\mu\nu} , 
  \quad
 b_{\mu\nu} := \partial_\mu b_\nu - \partial_\nu b_\mu .
  \quad
  \tilde \chi_{\mu\nu}= {1 \over 2}
\epsilon_{\mu\nu\alpha\beta}
 (\partial^\alpha \chi^\beta - \partial^\alpha \chi^\beta) .
 \label{Hd}
\end{eqnarray}
The tensor $B_{\mu\nu}$ has six degrees of freedom, while
the fields $b_\mu$ and $\chi_\mu$ have eight.  This mismatch
is not a problem, since two degrees are redundant; the
gauge transformation 
$
  b_\mu(x) \rightarrow b_\mu'(x) = b_\mu(x) - \partial_\mu
\theta,
$
$
  \chi_\mu(x) \rightarrow \chi_\mu'(x) = \chi_\mu(x) -
\partial_\mu \varphi,
$
leave $B_{\mu\nu}$ invariant.
In the function integral, the integration over $B_{\mu\nu}$
is replaced by an integration over $b_\mu$ and $\chi_\mu$,
provided that the gauge degrees of freedom are fixed, 
\begin{eqnarray}
 [dB_{\mu\nu}] = [db_\mu][d\chi_\mu] \delta(F[b]) \delta(F[\chi]) .
\end{eqnarray}
At one-loop level, integration over the redundant variable $\chi$
leads to
\begin{eqnarray}
  S_{APET}[a,b,k]
=
  \int d^4x \left[
    - {Z_a \over 4g^2}  f_{\mu\nu} f^{\mu\nu}
 - {1+z_b \over 4} g^2 b_{\mu\nu} b^{\mu\nu}
 - z_c b_\mu k^\mu 
 \right]  ,
 \label{APEGT}
\end{eqnarray}
where we have defined the magnetic current,
\begin{eqnarray}
  k^\mu := \partial^\nu \tilde f_{\mu\nu},
\quad \tilde f_{\mu\nu} :=  {1 \over 2} 
\epsilon_{\mu\nu\rho\sigma}f^{\rho\sigma} .
\end{eqnarray}
The magnetic current is non zero under the singular gauge
transformation such that the gauge transformed vector potential
satisfies the MAG. 
The resulting APEGT is written in terms of 
1) abelian gauge field $a_\mu$, 
2) the dual abelian gauge field $b_\mu$  
and 3) magnetic monopole current $k_\mu$ which couples to $b_\mu$.
This theory has
$U(1)_e \times U(1)_m$ symmetry 
\begin{eqnarray}
 a_\mu &\rightarrow& a_\mu + \partial_\mu \omega \quad (U(1)_e),
 \\
 b_\mu &\rightarrow& b_\mu + \partial_\mu \theta \quad (U(1)_m),
\end{eqnarray}
where $U(1)_m$ symmetry
is guaranteed by the topological conservation 
$
 \partial_\mu k^\mu = 0 .
$
The APEGT is considered as a quantum field theoretical realization of
't Hooft idea.

\section{Dual effective Abelian gauge theory}
\par
In order to obtain the dual theory with monopole condensation
which leads to the dual Meissner effect and quark confinement, we
consider the theory written in terms of $b_\mu$ alone. This can be
done as follows. It was pointed out that the magnetic monopole in
APEGT can be calculated from the original YM theory using the
off-diagonal gluons. In fact, we can show that the U(1)-invariant
current
$
  K^\mu := {1 \over 2}  
  \epsilon^{\mu\nu\rho\sigma} \partial_\nu (
 \epsilon^{ab3} A_\rho^a A_\sigma^b))   
 = {1 \over 2}  
  \epsilon^{\mu\nu\rho\sigma} \partial_\nu C_{\rho\sigma}
$
gives the magnetic monopole part of the abelian magnetic current
$k_\mu$ which contains both the magnetic monopole part and the Dirac
string part.
It turns out that the magnetic monopole charge calculated from this
definition satisfies the Dirac quantization condition.
The existence of magnetic monopole is a consequence of mathematical
identity,
$
  \Pi_2(SU(2)/U(1)) = \Pi_1(U(1)) = Z.
$

\par
By extracting the
$b_\mu$ dependent pieces from the action  (\ref{apBFYM3}), 
\begin{eqnarray}
S_{YM} = S_{YM}|_{b_\mu=0} +
  \int d^4x \left[ - {1 \over 4} g^2 b_{\mu\nu} b^{\mu\nu}
  + b^\mu K_\mu \right],
\end{eqnarray}
and  
inserting the identity
$
  1  =  \int [dK^\mu] \delta(K^\mu - {1 \over 2}  
  \epsilon^{\mu\nu\rho\sigma} \partial_\nu (
 \epsilon^{ab3} A_\rho^a A_\sigma^b)) , 
$
the partition function $Z_{YM}$ is written as
\begin{eqnarray}
  Z_{YM}[J] &:=& \int d\mu e^{-S_{YM}}
= \int [db_\mu] \exp \left\{  -
  \int d^4x \left[ - {1 \over 4} g^2 b_{\mu\nu} b^{\mu\nu}
  \right] \right\} 
 \nonumber\\&&
  \times \int d\tilde \mu \int [dK^\mu]
  \delta(K^\mu - {1 \over 2}   
  \epsilon^{\mu\nu\rho\sigma} \partial_\nu (
 \epsilon^{ab3} A_\rho^a A_\sigma^b)) 
 \nonumber\\&&
  \times \exp \left\{ -S_{YM}|_{b_\mu=0} -
  \int d^4x   
    b^\mu K_\mu    \right\} ,
\end{eqnarray}
where  $d\mu$ denotes the integration measure over 
all the fields.
\par
The dual effective theory with an action $S[b]$ is obtained by
integrating out all the fields except for $b_\mu$,
\begin{eqnarray}
  Z_{YM}[J] &:=& \int [db_\mu] 
  \exp \left\{ - S[b]  \right\} ,
\\   S[b] &=& {-1 \over 4}g^2 \int d^4x b_{\mu\nu} b^{\mu\nu}
+   \ln \langle \exp [\int d^4x b_\mu(x) K_\mu(x) ]
\rangle_0 ,
\end{eqnarray}
where the expectation value for a function $f$ of the field is
defined by
\begin{eqnarray}
 \langle  f(A)  \rangle_0 &:=&
\int d\tilde \mu \int [dK^\mu]
  \delta(K^\mu - {1 \over 2}  
  \epsilon^{\mu\nu\rho\sigma} \partial_\nu (
 \epsilon^{ab3} A_\rho^a A_\sigma^b)) 
 \nonumber\\&& \times
  \exp \left\{ -S_{YM}|_{b_\mu=0} \right\} 
f(A),
\end{eqnarray}
where $d\tilde \mu$ denotes the normalized measure without
$[db_\mu]$ so that
$
 \langle 1 \rangle_0 = 1.
$
It turns out that the cumulant expansion leads to
\begin{eqnarray}
   S[b] &=& {-1 \over 4}  g^2
   \int d^4x b_{\mu\nu}(x) b^{\mu\nu}(x)
+  \int d^4x \langle K_\mu(x) \rangle_0  b^\mu(x)
\nonumber\\&&
+ {1 \over 2} \int d^4x \int d^4y
\langle K_\mu(x) K_\nu(y) \rangle_c b^\mu(x) b^\nu(y) 
+ O(b^3) ,
\label{dualaction}
\end{eqnarray}
where $\langle K_\mu(x) K_\nu(y) \rangle_c$ is the
connected correlation function obtained from the normalized
expectation value.
\par
We can obtain a similar expression for the APEGT
using the action (\ref{APEGT}).  Hence the argument in
the next subsection can be extended also to the APEGT.

\section{Dual Meissner effect due to monopole condensation}

Recall that the effective dual abelian theory $S[b]$ has $U(1)_m$
symmetry. 
The magnetic current
 satisfies the conservation
$
 \partial_\mu K^\mu = 0 ,
$
consistently with the $U(1)_m$ symmetry.  
\par
When $U(1)_m$ is unbroken,  
the correlation function of the magnetic monopole current
is transverse,
\begin{eqnarray}
  \langle K_\mu(x) K_\nu(y) \rangle_c 
=  \left( \delta_{\mu\nu} \partial^2 
- {\partial_\mu \partial_\nu} \right)  M(x-y)    .
\label{kk}
\end{eqnarray}
As long as the magnetic $U(1)_m$ symmetry is not broken, 
the dual gauge field $b_\mu$ is always massless as can
be seen from (\ref{dualaction}) and (\ref{kk}).   Therefore
non-zero mass for the dual gauge field implies  breakdown
of the
$U(1)_m$ symmetry.
\par
If $U(1)_m$ symmetry is broken in such a way that
\begin{eqnarray}
  \langle K_\mu(x) K_\nu(y) \rangle_c 
=   g^2 \delta_{\mu\nu} \delta^{(4)}(x-y) f(x)
+ \cdots ,
\end{eqnarray}
 the mass term is generated,
\begin{eqnarray}
  S[b] = \int d^4x \left[ 
 {-1 \over 4} g^2  b_{\mu\nu}(x) b^{\mu\nu}(x)
  +  {1 \over 2} g^2 m_b^2 b_\mu(x) b_\mu(x) 
  + \cdots \right] ,
  \label{actionb}
\end{eqnarray}
if we write $f(x) = m_b^2$.
This can be called the  {\it dual Meissner effect};
the dual gauge field acquires a mass given
by
\begin{eqnarray}
  m_b^2 = {1 \over 4g^2} \Phi(0) ,
\end{eqnarray}
if the monopole  loop  condensation
occurs in the sense that,
\begin{eqnarray}
 \Phi(x) := \lim_{y \rightarrow x} 
  {\langle K_\mu(x) K_\mu(y) \rangle_c \over 
  \delta^{(4)}(x-y)}
\not= 0 .
\label{op}
\end{eqnarray}
This is a criterion of dual superconductivity of QCD.
In the translation invariant theory, $\Phi(x)$
is an $x$-independent constant which depends only on the
gauge coupling constant $g$.
\par
It should be remarked that
$\Phi$ is not the local order parameter in the usual
sense.  In order to find the non-zero value of
$m_b$, we must extract,
from the magnetic monopole current correlation
function
$
\langle K_\mu(x) K_\nu(y) \rangle_c ,
$
a piece which is proportional to
the Dirac delta function $\delta^{(4)}(x-y)$ diverging as
$y
\rightarrow x$.
Therefore, if such type of strong short-range correlation
between two magnetic monopole loops does not exist, $\Phi$
is obviously zero.  This observation seems to be consistent
with the result of lattice simulations.
The monopole loops exist both in the confinement and the
deconfinement phases.
However, in the deconfinement phase
the monopole currents are dilute and the vacuum contains
only short monopole loops with some non-zero density.
In the confinement phase, on the other hand, the monopole
trajectories form the infinite long loops and the monopole
currents form a dense cluster, although there is a number
of small mutually disjoint clusters. 
\par
Is this dual theory the same as the dual GL theory?
It should be remarked that APEGT doesn't need any scalar
field.  In this sense, the mechanism in which the dual
gauge field acquires a mass is different from the dual
Higgs mechanism.  Nevertheless, we can always introduce the
scalar field into APEGT so as to recover the spontaneously
broken
$U(1)_m$ symmetry,
\begin{eqnarray}
   {1 \over 2} m_b^2 b_\mu(x) b_\mu(x) 
   \rightarrow {1 \over 2} m_b^2 
   (b_\mu(x) -  \partial_\mu \theta (x))^2
   =   | (\partial_\mu - ib_\mu(x)) \phi(x)|^2 ,
\end{eqnarray}
where we identify 
\begin{eqnarray}
   \phi(x) = {m_b \over \sqrt{2}} e^{i\theta(x)} .
   \label{rfs}
\end{eqnarray}
Indeed, the  result is invariant under 
$ b_\mu \rightarrow b_\mu + \partial_\mu \alpha$ and
$ \theta \rightarrow \theta + \alpha$
($ \phi \rightarrow e^{i\alpha} \phi $).
Such a scalar field is called the
St\"uckelberg field or Batalin-Fradkin field.
The case (\ref{rfs}) is obtained as an extreme type
II limit (London limit),
\begin{eqnarray}
   \lim_{\lambda \rightarrow \infty} V(\phi), \quad
    V(\phi) := \lambda (|\phi(x)|^2 - m_b^2/2)^2 ,
\end{eqnarray}
or non-linear $\sigma$ model with a constraint in the integration
measure,
\begin{eqnarray}
 \delta(|\phi(x)|^2 - m_b^2/2) .
\end{eqnarray}

The value $\phi_0$ at which the potential $V(\phi)$ has a
minimum is proportional to the mass $m_b$ of dual gauge
field,
\begin{eqnarray}
  m_b = \sqrt{2} \phi_0 =  \sqrt{\Phi}/(2g) .
\end{eqnarray}
In the deconfinement phase, the minimum is given by
$\phi_0=0$ ($m_b=0$), while in the
confinement phase the minimum is shifted from zero
$\phi_0\not=0$ ($m_b\not=0$) which corresponds to
monopole condensation.  Thus the dual abelian gauge theory
with an action
$S[b]$ is equivalent to (the London limit of) the dual GL
theory (or the dual Abelian Higgs model with radial part of
the Higgs field being frozen), 
\begin{eqnarray}
  S[b] = \int d^4x \left[ 
 {-1 \over 4} b_{\mu\nu} b^{\mu\nu}
  +  | (\partial_\mu - ig^{-1} b_\mu ) \phi|^2
  + \lambda (|\phi|^2 - \phi_0^2)^2 + \cdots \right] ,
\end{eqnarray}
where we have rescaled the field 
$b_\mu \rightarrow b_\mu/g$.  Note that the inverse
coupling $g^{-1}$ has appeared as a coupling constant.  
This implies that the dual theory is suitable for
describing the strong coupling region.
Thus the dual GL theory is derived from YM theory without any
specific assumption.

\section{Dual description of low-energy physics}

Regarding the APEGT  with an action
$S[a,b,k]$ as an interpolating theory, we can obtain the dual
description of the same physics, say quark confinement, based on
$S[a], S[b]$. The respective effective theory is obtained by
integrating out the remaining field variables.
It is valid as an low-energy effective (abelian) gauge theory of
QCD in the length scale $R>R_c:=m_{A}^{-1}$, since the off-diagonal
parts with mass of order $m_A$ are integrated out.
\par
\vskip 0.5cm
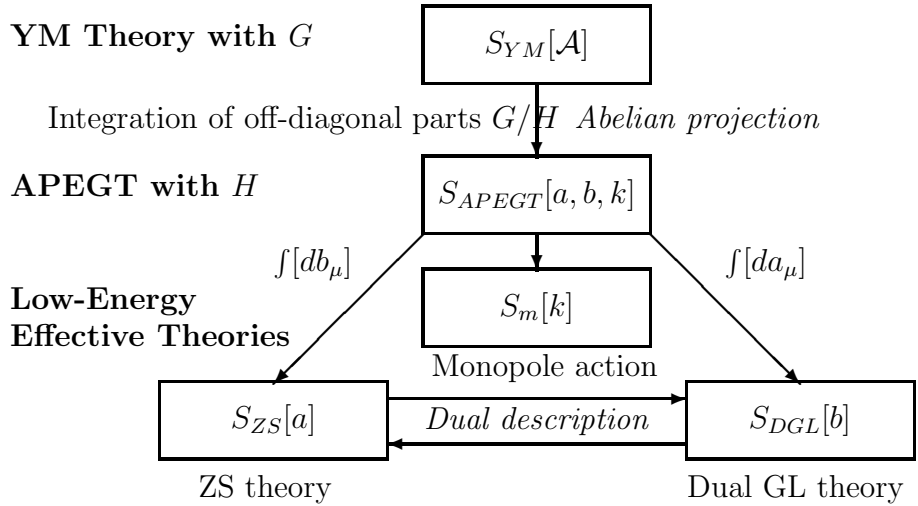
\begin{center}
\unitlength=1cm
\thicklines
 \begin{picture}(12,6)
 \put(5.5,5){\framebox(3,1){$S_{YM}[{\cal A}]$}}
 \put(7,5){\vector(0,-1){1}}
 \put(0,5.5){\bf YM Theory with $G$}
 \put(7.5,4.4){\it Abelian projection}
 \put(0.5,4.4){Integration of off-diagonal parts $G/H$}
 \put(5.5,3){\framebox(3,1){$S_{APEGT}[a,b,k]$}}
 \put(7,3){\vector(0,-1){0.5}}
 \put(0,3.5){\bf APEGT with $H$}
 \put(5.5,1.5){\framebox(3,1){$S_{m}[k]$}}
 \put(5.6,1.1){Monopole action}
 \put(5.5,3){\vector(-1,-1){2}}
 \put(3.5,2.5){$\int[db_\mu]$}
 \put(0,2){\bf Low-Energy}
 \put(0,1.5){\bf Effective Theories}
 \put(2,0){\framebox(3,1){$S_{ZS}[a]$}}
 \put(8.5,3){\vector(1,-1){2}}
 \put(9.5,2.5){$\int[da_\mu]$}
 \put(9,0){\framebox(3,1){$S_{DGL}[b]$}}
 \put(5,0.8){\vector(1,0){4}}
 \put(9,0.2){\vector(-1,0){4}}
 \put(5.5,0.4){\it Dual description}
 \put(2.5,-0.5){ZS theory}
 \put(9,-0.5){Dual GL theory}
 \end{picture}
\end{center}

\vskip 1cm
\par
(a) Dual GL theory $S_{DGL}[b]$ (non asymptotic-free abelian gauge
theory)
\begin{eqnarray}
  S_{DGL}[b] = \int d^4x \left[ 
 {-1 \over 4} b_{\mu\nu} b^{\mu\nu}
  +  | (\partial_\mu - ig^{-1} b_\mu ) \phi|^2
  + \lambda (|\phi|^2 - \phi_0^2)^2 + \cdots \right] .
   \label{actionbb}
\end{eqnarray}
This dual GL theory exhibits dual superconductivity, since it has
Nielsen-Olesen vortex solution corresponding to the Abrikosov vortex
in the GL theory describing the usual superconductivity.
When the monopole condensation occurs, $S[b]$ shows dual Meissner
effect $m_b=\sqrt{2}\phi_0\not=0$.  This leads to the linear static
potential. Therefore we obtain the dual superconductor vacuum of QCD.
\par
(b) Zwanziger-Suzuki (ZS) theory $S[a]$ (asymptotic-free abelian
gauge theory)   
The effective abelian theory written in terms of
$a_\mu$ is obtained 
(Using the Zwanziger formalism, this theory was first derived by
Suzuki
\cite{Suzuki88} assuming the abelian dominance, i.e. neglecting the
off-diagonal gluons from the beginning),
\begin{eqnarray}
 S_{ZS}[a] &=& \int d^4x \left[
 {-1 \over 4g(\mu)^2}  f_{\mu\nu}(x) f^{\mu\nu}(x)
   + {1 \over 2} a^\mu(x) {n^2 m_b^2 \over 
   (n \cdot \partial)^2+n^2 m_b^2} X_{\mu\nu}
(\partial)  a^\nu(x) \right] ,
   \nonumber\\
   && X_{\mu\nu}(\partial) := {1 \over n^2} 
   \epsilon^{\lambda\mu\alpha\beta}
   \epsilon^{\lambda\nu\gamma\delta} n_\alpha n_\gamma
   \partial_\beta \partial_\delta ,
   \label{actiona}
\end{eqnarray}
where $n$ is an arbitrary fixed four-vector appearing in
the Zwanziger formalism.  The coupling constant $g(\mu)$ is
the  running coupling constant obeying the same $\beta$
function as the YM theory. 
Note that the local $U(1)_e$ symmetry is not broken and $a_\mu$ is
massless, since 
$
 \partial^\mu X_{\mu\nu} \equiv 0 \equiv \partial^\nu X_{\mu\nu} .
$
\par
Both effective theories (\ref{actionbb}), (\ref{actiona})
leads to the linear static potential with the string tension
$\sigma$, 
\begin{eqnarray}
 V(r) = \sigma r ,
 \quad \sigma = {Q^2 \over 4\pi} m_b^2 f(\kappa_{GL}) ,
\end{eqnarray}
where $f(x)$ is a dimensionless function.
The essential part
$m_b^2$ in the string tension follows simply due to the
dimensional analysis, irrespective of the details of the
calculation. 
\par
(c) Monopole action $S[k]$ (Lattice version was proposed by Smit and
van der Sijs\cite{SS91})
\begin{eqnarray}
  S_m[k]  =  \int d^4 x \left[
    - {1 \over 4g^2(\mu)}   f_{\mu\nu}(x) f^{\mu\nu}(x)
 + {1 \over g^2} 
 k^\mu(x) D_{\mu\nu}(x-y) k^\nu(y) \right]  ,
 \label{mlaction}
\end{eqnarray}
where $D_{\mu\nu}(x-y)$ is the massless vector propagator.
By using the monopole action and energy(action)-entropy argument, 
we can show that in the strong coupling region long monopole loops
give dominant contribution to the path integral and that the
non-zero monopole condensation is obtained.

\section{Addenda}
The above strategy for obtaining APEGT can be extended to $SU(N)$ YM
theory, see Ref.~\citen{Kondo98b}.
\par
In this report, we have not discussed the low-energy effective
theory from a viewpoint of topological field theory.
By pushing ahead the idea of topological field theory, 
quark confinement has been proven recently, see Ref.~\citen{Kondo98} 
where massiveness of the off-diagonal gluons is also proved. 
The mass $m_A$ is essentially the same as the Haldane gap in
one-dimensional quantum Heisenberg antiferromagnets.

\section*{Acknowledgements}
I would like to thank organizers of YKIS'97 for inviting me to give
a talk.



\end{document}